\documentstyle[preprint,prb,aps]{revtex}
\begin{document}
\draft
\title{Emission Spectrum of a Dipole in a Semi--infinite Periodic Dielectric Structure: Effect of the Boundary}

\author{A.~G.~Galstyan,  M.~E.~Raikh  and Z.~V.~Vardeny}
\address{Department of Physics, University of Utah, 
Salt Lake City, Utah  84112}

\maketitle
\begin{abstract}
The  emission spectrum of a dipole embedded in a semi--infinite photonic crystal is calculated. For simplicity we study the case in which the dielectric function is sinusoidally modulated only along the direction perpendicular to the boundary surface plane. In addition to oscillations of the emission rate with the distance of the dipole from the interface  we also observed that the shape of the emission spectrum srongly depends  on the $\em initial$ $\em phase$ of the dielectric modulation. When the direction of light propagation inside the periodic structure is not normal to the boundary surface plane we observed aditional singularities in the emission spectrum, which arise due to different angle--dependence of the Bragg stop--band for $TE$ and $TM$ polarizations.
\end{abstract}
\narrowtext

\section{Introduction}
\label{Intro}
It is well known that fluorescence lifiteme of  an atom can be drastically changed when placed in an inhomogenous medium. In his pioneering work Purcell\cite{purcell} predicted a strong enhancement of the radiative decay rate of an  emitter placed inside a resonant microcavity. In contrast, it was also suggested\cite{kleppner} that spontaneous emission can be totally inhibited if the emitter transition frequency lies below that of the fundamental resonator mode. This effect  can be understood in terms of redistribution of photonic density of states (DOS) caused by inhomogeneity and/or nontrivial boundary conditions imposed on the radiative field. During the years the spontaneous emission of a dipole coupled to various optical enviroments such as metallic cavities\cite{goy}, Fabry Perot two--mirror cavities\cite{chance}, dielectric microspheres\cite{chew} and nanobubbles,\cite{klimov} has been a subject of extensive theoretical and experimental studies.\cite{haroche}

Recently there has been a growing interest in studies of radiative properties of fluorescent molecules inside  periodic dielectric structures, so called photonic crystals (PC).\cite{yablonovitch,john87} The Bragg diffraction of light that occurs in these systems opens up a spectral  gap\cite{soukoulis} (or a pseudo--gap) in the photonic DOS in analogy with electronic energy gaps in semiconductors. A quantum electro--dynamical approach for the radiative decay inside PC has revealed novel physical phenomena such as strong suppression of spontaneous emission within the gap and formation of photon--atom bound states,\cite{john91,john94,rupasov96,john90}  localization of superradiant modes near the band edges,\cite{john95} $\em etc.$. Calculations of emission spectra  within the framework of classical theory were  performed for  one--dimensional Kronig-Penny type model\cite{dowling} as well as for three--dimensional fcc lattice structures\cite{suzuki}. It has been established that  inhibition of spontaneous emission within the gap is  accompanied with strong enhancement at the band edges.\cite{dowling,suzuki} It was also shown that the emission spectrum strongly depends on the position of the emiter within the unit cell,\cite{dowling,suzuki} as well as on its orientation.\cite{suzuki} Experimental observations of  inhibited spontaneous emission have been reported for different periodic structures.\cite{martorell,tong,yamasaki,petrov,yoshino,megens}

So far, the exsisting theoretical studies have considered infinite periodic structures. As a result, the emission power  was identically zero within the gap.\cite{dowling,suzuki} In the experiments,\cite{martorell,tong,yamasaki,petrov,yoshino,megens} however,  the photonic gap appears as a drop by a factor of $\sim2$ in the emission power within a certain spectral interval. This points at the important role of the boundary between the photonic crystal and the air, which is studied  in the present paper. As we will show below, accounting for a nearby interface results in a nontrivial dependence of the emission spectrum on the $\em initial$ $\em phase$ of the dielectric modulation. If the distance of the dipole from the boundary surface plane and the dielectric modulation period are respectively $d$ and $a$, then one might expect that the dependence of the emission spectrum on the $\em initial$ $\em phase$  be small in the parameter $a/d$. On the contrary, we found that the strong dependence of the emission spectrum on the $\em initial$ $\em phase$ persists even in the limit $d/a \rightarrow\infty$ (provided there is no absorption in the system). We illustrate this effect in the frame of the simplest possible model. Namely, we choose the dielectric modulation to be $\em i$) weak, $\em ii$) one--dimensional, $\em iii$) sinusoidal. To quantitavely study the effect of a plane boundary  we generalize the standard calculations of the emission rate in periodic media for the case of semi--infinite geometry (Sec. 2). In Sec. 3 we present numerical results for emission spectra  illustrating the role of the $\em initial$ $\em phase$. Discussion of our results and their relevance to recent measurements is presented in Sec. 4.

\section{Derivation of the Power Emission Spectrum}
We  schematically depict the system under consideration in Fig. 1. The dielectric function for the left half--space is constant and equals $\varepsilon_0$, whereas for $x>0$  is given by
\begin{equation}
\label{epsx}
\varepsilon (x) = \varepsilon _{1}+\delta\varepsilon \cos(\sigma x +\phi) \ , 
\end{equation}
where $\phi$ is the initial phase of the dielectric modulation, $\sigma=2\pi/a$ is the modulation wave vector, and $\delta \varepsilon$ is the amplitude of the modulation.  Below we assume $\delta \varepsilon \ll \varepsilon_1$. The wave  equations for the elctric and magnetic fields  are
\begin{equation}
\label{WE}
{\bf \nabla}^2 {\bf E({\bf r})}-{\bf \nabla}({\bf \nabla} {\bf E({\bf r})})   +\frac{\omega^2}{c^2}\varepsilon(x){\bf E({\bf r})}=i\frac{4\pi \omega}{c^2}{\bf J({\bf r})} 
\end{equation} 
\begin{equation}
\label{WB}
{\bf \nabla}^2 {\bf B({\bf r})}-{\bf \nabla}[\ln\varepsilon(x)]\times {\bf \nabla}\times {\bf B({\bf r})}   +\frac{\omega^2}{c^2}\varepsilon(x){\bf B({\bf r})}=-\frac{4\pi}{c}{\bf \nabla}\times  {\bf J({\bf r})} 
\end{equation} 
where the  radiation source ${\bf J} ({\bf r} )={\bf j} \delta({\bf r}-{\bf r_0})$   is located at point ${\bf r_0}=(d,0,0)$. We note that the term proportional to $\delta \varepsilon$ has been neglected in the r.h.s.  of Eq. (\ref{WB}). The time averaged radiative power per unit solid angle is given by
\begin{equation}
\label{power1}
\frac{dP}{d \Omega}=\frac{c}{8 \pi} Re \biggl [r^2 {\bf n}({\bf E} \times {\bf B^{*}} ) \biggr ],
\end{equation}
where $B^{*}$ is the complex conjugate of $B$, $r=|{\bf r}|$ and ${\bf n}={\bf r}/r$ is the unit radius vector. Without any loss of generality we choose $n_z =0$ (see Fig. 1). Then it is  very convenient to separately treat two possible orientations of the dipole. Indeed, one can easily check that the current density  components $J_z$ and $J_{x}, J_y$ give rise to Electro--Magnetic ($EM$) radiation with respectively electric ($TE$ polarization) and magnetic ($TM$ polarization) fields polarized in the $z$ direction. Since these two modes do not interfere, their  contributions to the radiation power are additive. The corresponding $EM$ wave equations for the two polarizations are obtained from the $z$--components of Eqs. (\ref{WE}) and (\ref{WB}) by taking the Fourier transform with respect to $y$ and $z$ coordinates:
\begin{equation}
\label{Ez}
\frac{d^2 E_z }{dx^2}+\biggl(\frac{\omega ^2}{c^2}\varepsilon(x)-k_y^2 \biggr )E_z= \frac{4 \pi i \omega}{c^2}j_z \delta(x-d) \ , \ (TE)
\end{equation}
\begin{equation}
\label{Bz}
\frac{d^2 B_z }{dx^2}-\biggl ( \frac{\partial \ln \varepsilon }{\partial x }\biggr ) \frac{d B_z }{dx} + \biggl(\frac{\omega ^2}{c^2}\varepsilon(x)-k_y^2 \biggr )B_z= -  \frac{4 \pi}{c}\biggl (\frac{\partial j_y}{\partial x}-ik_y j_x \biggr )\delta(x-d) \ , \ (TM)
\end{equation}
where $k_y$ is the $y$--component of the wave vector ($E_z(x;k_y), B_z(x;k_y) \propto e^{i k_y y}$). Since we want to calculate the power emmited in $xy$ plane we have set $k_z=0$ in Eqs. (\ref{Ez}) and ($\ref{Bz})$. The solution of the corresponding homogenous equations may be written as  a sum of incident, reflected and transmitted $EM$ waves, with  two linearly independent terms $E_1(x)$, $E_2(x)$ and $B_1(x)$, $B_2(x)$  corresponding to the incident $EM$ wave from the right and left, respectively (see inset of Fig. 1). To solve  Eqs. (\ref{Ez}--\ref{Bz})  we employ the  variation of a constant method.  We seek solution in the form:
\begin{equation}
\label{CV}
E_z(x)=C^{E}_1(x)E_1(x)+C^{E}_2(x)E_2(x) \ , \ B_z(x)=C^{B}_1(x)B_1(x)+C^{B}_2(x)B_2(x)
\end{equation} 
Upon substituting (\ref{CV}) into (\ref{Ez}--\ref{Bz}) we find for the variational coefficients
\begin{equation}
\label{c1c2}
C^{E}_{1,2}(x)=  i \omega  \frac{4 \pi}{c^2}  W_{E}^{-1}\int _{X_{1,2}^{E}}^{x} dx^{\prime}  E_{2,1}(x^{\prime}) j_z \delta(x^{\prime}-d)  
\end{equation}
\begin{equation}
\label{c1c2tld}
C^{B}_{1,2}(x)= - \frac{4 \pi}{c} W_{B}^{-1}\int _{X^{B}_{1,2}}^{x} dx^{\prime} B_{2,1}(x^{\prime}) \biggl( \frac{\partial j_y}{\partial x^{\prime}} - ik_y j_x \biggr) \delta(x^{\prime} -d)  
\end{equation}
where $W_{E}$, $W_{B}$  are the Wronskians
\begin{equation}
\label{wronskian}
W_{E}=E_1 \frac{dE_2}{dx}-E_2 \frac{dE_1}{dx} \ , \ W_{B}=B_1 \frac{dB_2}{dx}-B_2 \frac{dB_1}{dx}
\end{equation}
and  $X^{E}_{1,2}$, $X^{B}_{1,2}$ in the lower integration limits are constants of integration. They  are  determined from the boundary conditions that there are no incoming $EM$ waves, implying $X^{E}_1= X^{B}_1 = \infty$, $X^{E}_2 = X^{B}_2 =-\infty$.  
Hence, the solutions of Eqs. (\ref{Ez}--\ref{Bz}) for large negative $x$ which satisfy the boundary conditions can be written as follows
\begin{equation}
\label{SolEz}
E_z(x)= i \omega  \frac{4 \pi}{c^2} W_{E}^{-1} j_z E_1(d)E_2(x) \ ,
\end{equation}
\begin{equation}
\label{SolBz}
B_z(x)= \frac{4 \pi}{c} W_{B}^{-1} \biggl ( j_y\frac{dB_1}{dx}\biggl|_{x=d} +  ik_yj_xB_1(d)\biggr )B_2(x) \ .
\end{equation}
It can be easily shown that for negative $x$ one has $W_{E}^{-1}E_2(x)=W_{B}^{-1}B_2(x)=ie^{-ik_x x}/2k_x$. Thus, the remaining task is to find $E_1(d)$, $B_1(d)$ and $(dB_1/dx)_{x=d}$.

The solution of the homogeneous equations for the left  half--space can be written as a sum of two plane waves,
\begin{equation}
\label{plwaves}
E_1(x) = e^{-i k_x x} + R_{E} e^{i k_x x} \ , \ B_1(x) = e^{-i k_x x} + R_{B}e^{i k_x x} 
\end{equation}
where $R_{E}$, $R_{B}$ are the optical reflection coefficients for $TE$ and $TM$ polarizations, respectively. For the right  half--space one may use the Bloch theorem to find the solution
\begin{equation}
\label{bloch}
E_1(x)=e^{ i q_E x} \sum _{n=-\infty}^{\infty} A^E_n e^{i \sigma n x} \ , \
B_1(x)=e^{ i q_B x} \sum _{n=-\infty}^{\infty} A^B_n e^{i \sigma n x} \ . 
\end{equation}
When substituting Eq.(\ref{bloch}) into Eqs.(\ref{Ez}--\ref{Bz}) we obtain two  infinite systems of linear, homogenous  equations for the coefficents $A^E_n$, $A^B_n$
\begin{equation}
\label{linsys1}
\biggl ( \frac{\omega ^2}{c^2}\varepsilon_1 -k_y^2-(q_E+n \sigma)^2 \biggr )A^E_n +\frac{\omega ^2}{c^2} \frac{\delta \varepsilon}{2} \biggl ( e^{i \phi}A^E_{n-1}+ e^{-i \phi}A^E_{n+1} \biggr ) = 0 
\end{equation}
\begin{eqnarray}
\label{linsys2}
&& \biggl ( \frac{\omega ^2}{c^2}\varepsilon_1 -k_y^2-(q_B+n \sigma)^2 \biggr )A^B_n + \frac{\delta \varepsilon}{2 \varepsilon_1} e^{i \phi} \biggl(\frac{\omega ^2}{c^2}\varepsilon_1+\sigma(q_B+(n-1)\sigma)\biggr) A^B_{n-1}  \nonumber \\ 
&&~~~~~~~~~~~~~~~~~~~~~~~~~~~~~~~~~~~~+ \frac{\delta \varepsilon}{2 \varepsilon_1} e^{-i \phi} \biggl(\frac{\omega ^2}{c^2}\varepsilon_1-\sigma(q_B+(n+1)\sigma)\biggr) A^B_{n+1}=0 \ . \
\end{eqnarray}
We note that the $\em initial$ $\em phase$ $\phi$  explicitly enters into these equations.

Near the Bragg resonance that occurs at $q_{E,B} \approx \sigma /2$, the main coefficents which contribute to the sums in Eq.(\ref{bloch}) are $A^E_0$, $A^B_0$ and $A^E_{-1}$, $A^B_{-1}$ since the rest of $A^{E,B}_n$ are small in the parameter $\delta \varepsilon / \varepsilon_1$. In this approximation the equation systems (\ref{linsys1}--\ref{linsys2}) may be simplified into two $2 \times 2$ matrix equations. Requiring for the determinants of these matrices to vanish, one finds the dispersion relations for the two $EM$ polarizations near  resonance 
\begin{equation}
\label{dispte}
\delta q_E = \frac{\sigma}{2\kappa}\sqrt{ \biggl (\frac{\delta \omega}{\omega_0}\biggr )^2-\Delta^2} 
\end{equation}
\begin{equation}
\label{disptm}
\delta q_B = \frac{\sigma}{2\kappa}\sqrt{ \biggl (\frac{\delta \omega}{\omega_0}\biggr )^2-\Delta^2 (1-2\kappa)^2} \ , \
\end{equation}
where we have introduced $\Delta=\delta \varepsilon /4 \varepsilon_1$, $q_{E,B}-\sigma /2 =\delta q_{E,B} \ll \sigma /2$, $\omega - \omega_0 = \delta \omega \ll \omega_0 $, $\kappa = \sigma^2 c^2/4\omega_0^2 \epsilon_1$ and the $k_y$--dependent resonant frequency $\omega_0$ is given by
\begin{equation}
\omega_0 = \frac{c}{\varepsilon_1} \sqrt{\sigma^2 /4 + k_y^2} \ . \
\end{equation}
Eqs. (\ref{dispte}--\ref{disptm}) show that there is a  spectral gap for $EM$ waves propagating in the system centered at $\omega_0$.  For both $TE$ and $TM$  polarizations the central gap position shifts to higher frequencies with increasing  $k_y$ (which also determines the propagation direction  of the radiative field). For $TE$ polarization the gap broadens with increasing angle $\theta^{\prime}$ (see Fig. 1) whereas for $TM$ polarization the gap narrows and  disappears at $2\kappa=1$, which corressponds to the propagation direction for which $k_y=\sigma/2$. This can be defined as a Brewster angle for  Bragg diffraction. If one increases  $k_y$ further then the gap reopens again.

Using Eqs. (\ref{linsys1})--(\ref{disptm}) we obtain the following expressions for the electric and magnetic fields for the right half space:
\begin{equation}
\label{Efield}
E_1(x)=A_{-1}^Ee^{i\delta q_E x }\biggl ( e^{i \frac{\sigma}{2}x}-F_Ee^{-i\phi}e^{-i \frac{\sigma}{2}x}\biggr)
\end{equation}
\begin{equation}
\label{Bfield}
B_1(x)=A_{-1}^Be^{i\delta q_B x }\biggl ( e^{i \frac{\sigma}{2}x}-F_Be^{-i\phi}e^{-i \frac{\sigma}{2}x}\biggr)
\end{equation}
Here the functions $F_E$ and $F_B$ describe the coupling between incident and Bragg reflected waves and are defined as follows:
\begin{equation}
 F_{E}( \delta \omega, \theta)=\frac{1}{\Delta}\biggl (\frac{\delta \omega}{\omega_0} -\sqrt{ \biggl (\frac{\delta \omega}{\omega_0}\biggr)^2-\Delta^2} \ \biggr )  
\end{equation}
\begin{equation}
 F_{B}( \delta \omega , \theta)=\frac{1}{\Delta (1-2 \kappa)}  \biggl (\frac{\delta \omega}{\omega_0} -\sqrt{ \biggl (\frac{\delta \omega}{\omega_0}\biggr)^2-\Delta^2 (1-2\kappa)^2} \ \biggr ) \ . \
\end{equation}
Finally, we match the fields components and their derivatives  at $x=0$, take the inverse Fourier transform with respect to $k_y, k_z$  and use Eq. (\ref{power1}) to get the following expression for the total radiated power $dP/d\Omega$, normalized to the radiation power when there is no dielectric modulation:
\begin{eqnarray}
\label{power2}
&& \biggl (\frac{dP}{d\Omega}\biggr)_N =\frac{j_z^2 T_{0E}^2 }{j_z^2 T_{0E}^2 +\varepsilon_1 j_1^2 T_{0B}^2} \biggl|\frac{1-F_{E}e^{-i(\sigma d + \phi)}}{1-R_{0E} F_{E}e^{-i \phi}}e^{i \delta q_E d} \biggr |^2 \nonumber \\
&&~~~~~~~~+\frac{\varepsilon_1 j_1^2 T_{0B}^2 }{j_1^2 T_{0E}^2+\varepsilon_1 j_1^2 T_{0B}^2} \biggl|\frac{1-\chi F_{B}e^{-i(\sigma d + \phi)}}{1-R_{0B} F_Be^{-i \phi}} e^{i \delta q_B d} \biggr |^2 \ . \
\end{eqnarray}

Here $T_{0E}, R_{0E}$ and $T_{0B}, R_{0B} $ are the conventional Fresnel's  transmission and reflection coefficients from the dielectric interface without the dielectric modulation for the amplitudes of $TE$ and $TM$ polarized waves, respectively, $j_1=j_y \cos \theta^{\prime} +j_x \sin \theta^{\prime}$, where $\theta^{\prime}$ is related to  the observation angle $\theta$ by  Snell's law. The quantity $\chi$, which depends on the dipole orientation in the $xy$ plane is defined as  $\chi=(j_y-j_x \tan \theta^{\prime})/(j_y+j_x \tan \theta^{\prime})$.

Eq. (\ref{power2}) is the main result of this paper. We note that the phase $\phi$ is explicitly present in  both the numerator and denominator of Eq. (\ref{power2}), indicating the important role of the $\em initial$ $\em modulation$ $\em phase$ at the boundary interface.  In the next section we  numerically analyze the role of $\phi$ in the emission spectrum  for different situations.

\section{Numerical Results}

Before presenting our numerical results let us concentrate on a particular experimental case of  opals\cite{petrov,yoshino} and opal replica.\cite{yoshino} These artificial photonic crystals consist of closely packed $SiO_2$ spheres forming an fcc structure that contains fully interconnected voids. The opal replica  is formed by filling these voids with a precursor polymer solution  and then etching  the $SiO_2$ spheres after polymerization.\cite{yoshino} The opals and opal replica PC are  infiltrated with  various fluorescent  dye solutions to provide a radiation source inside the crystal. Inhibited spontaneous emission of the dye molecules in such  PC have been recently studied by several groups.\cite{petrov,yoshino} The refractive index  $n$ of $SiO_2$ is $n\approx 1.46$ and the refractive index contrast $\Delta n$ between the $SiO_2$ balls and the dye solution ranges between $\Delta n \approx 0.1-0.3$. This is not sufficient for a formation of a $\em complete$ photonic band gap. Instead, the system posseses  pseudogaps (or partial gaps) with an angle--dependent central frequency. To compare our results  to the experiments we have chosen in our model  $\Delta=\delta \varepsilon/4\varepsilon_1 = 0.1$.

First, we consider the case when the emitter is many periods away from the interface ($d=5a$).\cite{note} In Fig. 2 we show the emission power $\em averaged$ over the orientations of the emitter $\em as$ $\em well$ $\em as$ $\em over$ $\em its$  $\em position$ within the unit cell (the zero on the frequency axis in all plots corresponds to the central gap frequency $\Omega_0$ at $\theta=0$, where $\Omega_0=\sigma c/(2\sqrt{\varepsilon_1})$). Inside the gap, the emission power is strongly suppressed. It is seen, however,  that even in the limit of large $d/a$ the features of the $\em averaged$ emission power still depend on the $\em initial$ $\em phase$.

The evolution of $(dP/d\Omega)_N$ with $\phi$  outside the gap is described as follows. At $\phi=0$ there is a well pronounced singularity at the lower edge of the spectral gap. With increasing $\phi$, this singularity diminishes whereas another singularity starts to develop at the upper edge; the spectrum becomes symmetric at $\phi=\pi/2$. Further increase of $\phi$  leads to a gradual transformation of the initial curve into its mirror image with respect to the central gap frequency, the singularity now occuring at the upper band edge.

Fig. 2 corresponds to  $\theta =0$. The sensitivity of the emission power to $\phi$ appears to be even more pronounced for $\theta \neq 0$. We illustrate this effect by plotting in Fig. 3 the $\em averaged$ $(dP/d\Omega)_N$ for $\theta=60^o$  for the set of $\em initial$ $\em phases$ $\phi=0, \pi/4, \pi/2$. With increasing $\phi$ again, it is seen that there is a tendency for the emission spectrum to become symmetric near $\phi=\pi/2$. At $\phi=\pi$ (not presented here) the emission spectrum is again transformed into the mirror image of the initial curve at $\phi=0$ but now with respect to the shifted central gap frequency $\Omega_0/cos\theta^{\prime}$. We also note the appearance of  additional singularities in the emission spectrum. Their origin lies in the different angle--dependenceies of the gap width for $TE$ and $TM$ polarizations, as  seen before in Eqs (\ref{dispte}) and  (\ref{disptm}). To be more specific, we note that the highest and lowest frequency peaks correspond to the band edges for $TE$ polarization. Similarly, two peaks at the  intermidiate frequencies determine the band edges for $TM$ polarization. Both pairs of the singularities are located symetrically around the shifted Bragg frequency $\omega_0=\Omega_0/\cos\theta^{\prime}$.

Let us now turn to the discussion of the case where the emitter is close to the interface. The main feature of this situation is that $(dP/d\Omega)_N$ $\em inside$ the gap remains finite. One can see from Eq. (\ref{power2}) that moving the emitter $N$ periods away from the boundary decreases the emission power at the center of the gap by a factor of $exp(-2\pi N \Delta)$. Therefore, to study the features of the spontaneous emission inside the gap we choose $N=1,2$. This situation is illustrated in Fig. 4 where we plot the averaged $(dP/d\Omega)_N$  for the observation angle $\theta=60^o$. One can see that the evolution of the emission power with increasing the $\em initial$ $\em phase$ is very strong. Again, we note that there are four well pronounced singularities in the emission spectrum that corresponds to the spectral gap edges of the $TE$ and $TM$ polarizations. In particular, for $\phi=0$ there is a noticable enhancement of the emission rate (by a factor of $\sim 2$) at the frequency which determines the lower band--edge of the  $TE$ polarization, whereas for $\phi=\pi/2$ a similar enhancement occurs at the  edges of the $TM$ polarization gap. 

Finally, in Fig. 5 we plot the emission power integrated over the observation angle $\theta$ (i.e., the total power emitter in  $xy$ plane) for $\phi=0$ and $\phi=\pi$. In this case too an averaging over the dipole position within the unit cell has been performed. To allow for an unpolarized emission  we have chosen $j_x=j_y=j_z$. Remarkably, even after angular  averaging a weak dependence of the emission power on the initial phase still persists.

\section{Discussion}

We have studied the emission spectrum of a dipole inside a one--dimensional periodic  structure in the presence of a nearby plane boundary. As expected, the emission rates are strongly suppressed for  frequencies inside the spectral gap, provided that the emitter is many periods away from the interface ($d/a >5$). For frequencies near the band edges, the emission spectrum changes drastically with the $\em initial$ $\em phase$ of the dielectric modulation. We also observed  enhancement of the emission rates near the band edges, however, by a  factor much smaller than predicted in the previous studies.\cite{dowling,suzuki}  This can be attributed to the fact that the modulation in our model is weak. In Ref. \onlinecite{suzuki}, where the total radiation power from a dipole inside a infinite $3D$ fcc lattice was calculated numerically, the contrast in the dielectric constant $\Delta \varepsilon$ was more than $10$ allowing for the formation of a complete  photonic band gap. This resulted in an  enhancement of the radiated power at the band edges by a factor of $\sim 25$. Similarly, in Ref. \onlinecite{dowling} where the authors considered $1D$ Kronig--Penney type modulation of the refractive index, the enhancement factor at the band edges was about $\sim 30$,  whereas   inside the gap it was identically zero. Apparently, this resulted from  consideration of radiative modes polarized parallel to the dipole direction.  Allowing for nonpolarized radiation ($\em i. e.$ in all directions) should lead to qualitatively different results (see for example Fig. 5). Strictly speaking, the straightforward comparision of our calculation with those of previous studies is not possible due to the different approach developed here. However, it is clear from our consideration that the sensitivity of the power emission to the boundary conditions should persist also for strong and non--sinusoidal modulation and become even stronger.

In this paper we also studied  the case where the emitter is sufficiently close to the interface, so that the emission power for the frequencies inside the gap is finite. In this case also we showed that the features of the emission spectrum are very sensitive to the initial phase of the periodic modulation, hence emphasizing  the effect of the boundary. 

We note that in our calculations we assumed that no deffects exist in the system. In the presence of weak disorder the emission spectrum would be significantly modified. Let us consider radiation from a dipole that is many periods away from the boundary with frequency inside the spectral gap. If there are no defects  then the radiation ($e.$ $\em g.$, in the direction normal to the boundary) is strongly attenuated (see Fig. 2). However, introducing a small concentration of defects  opens up a new mechanism for the light emission to come out from the PC in the direction $\em normal$ to the boundary. Namely, light can propagate  without any attenuation in  directions for which the Bragg condition is not satisfied and then scatter off  defects that are close to the interface. As a result, the emission power in the direction normal to the boundary is not exponentially small but is finite, similar to the case of a dipole close to the boundary. Remarkably, only  defects close to the boundary, within the Bragg atenuation lenght $\xi_B=(2\pi\sigma \Delta)^{-1}$ from the interface, contribute to the emission power. Owing to this effect, we conclude that the emission spectrum in the presence of a weak disorder is rather universal, since it does not depend on the dipole--interface distance $d$. Experimentally this was demonstrated in Ref. \onlinecite{megens}, where the authors found a way to excite fluorescent dye molecules at different distances from the boundary. They have demonstrated that the emission power  at the center of the spectral gap is attenuated by a factor of $\sim 2$ and does not change upon changing $d$. This was interpretated in terms of light scattering off defects close to the PC interface.   

\section{Acknowledgements}
The authors thank N. Eradat, Drs. A. A. Zakhidov and R. H. Baughman for useful discussions. This work was supported in part by NSF under grant DMR--9732820 and the International Research Foundation of NEDO (Japan). One of the authors (M.~E.~R.) acknowledges the support of Petroleum Research Fund under grant ACS--PRF \#34302--AC6.

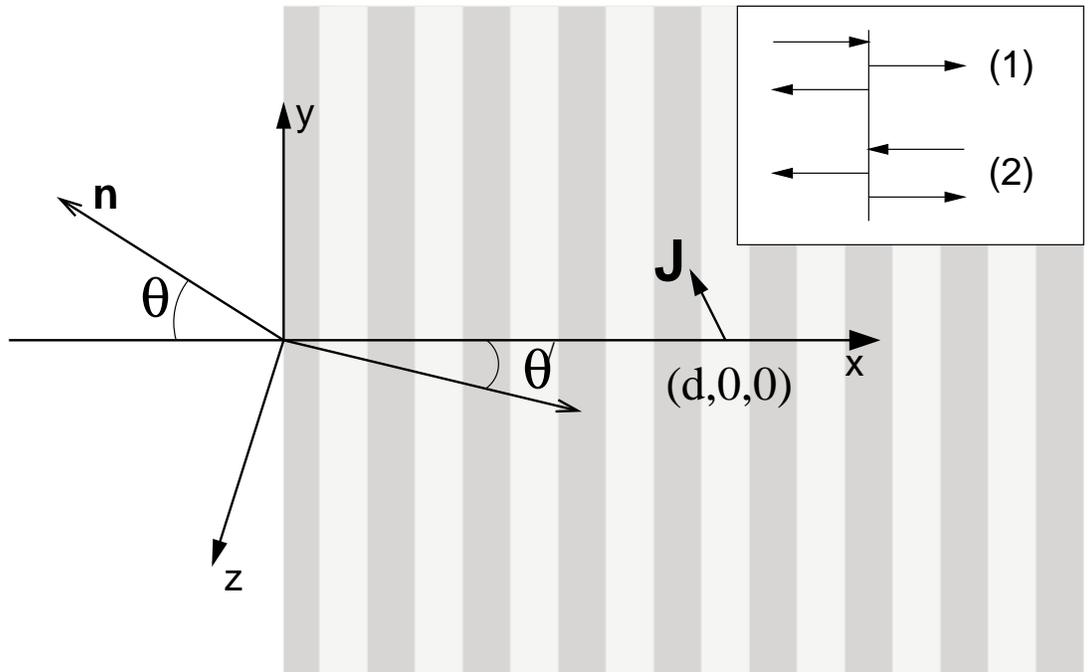
\begin{figure}
\caption{Schematic representation of our PC system. The emitter is located on the $x$ axis at a distance $d$ from the interface. The unit vector $\bf n$ lies in $xy$ plane. Inset: Illustration of two linearly independent solutions of the homogenous equations discussed in the text.}
\end{figure}

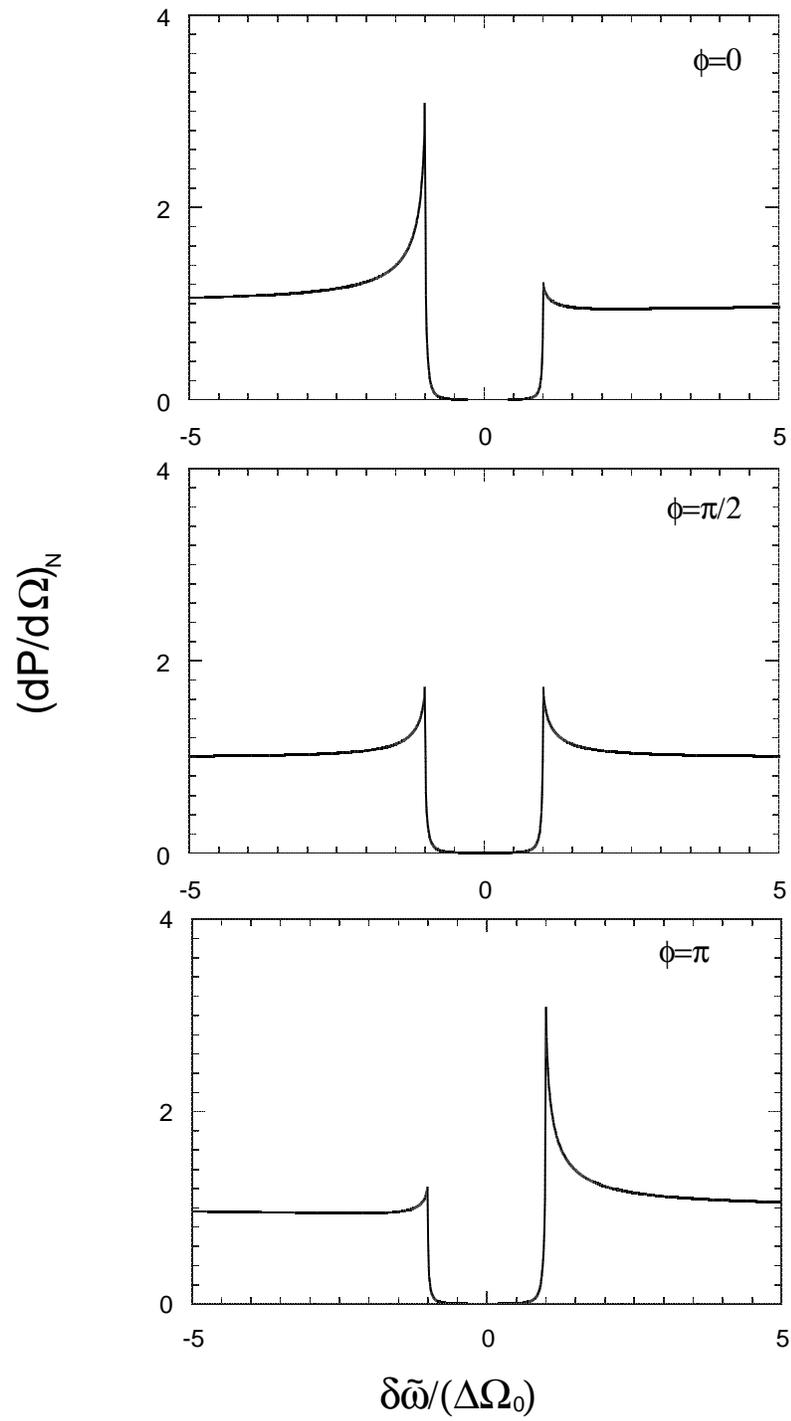
\begin{figure}
\caption{Normalized emission power  $(dP/d\Omega)_N$ averaged over the dipole emitter orientations at $\theta=0$ for $\phi=0, \pi/2$ and $\pi$. Here $\delta \tilde  \omega = \omega - \Omega_0$ is the deviation from the Bragg frequency at $\theta=0$. Calculations were performed with $\varepsilon_1=3\varepsilon_0=3$ and  $\Delta = 0.1$. The emitter is five periods away from the interface ($d=5a$) implying strong suppression of the emission within the gap.} 
\end{figure}

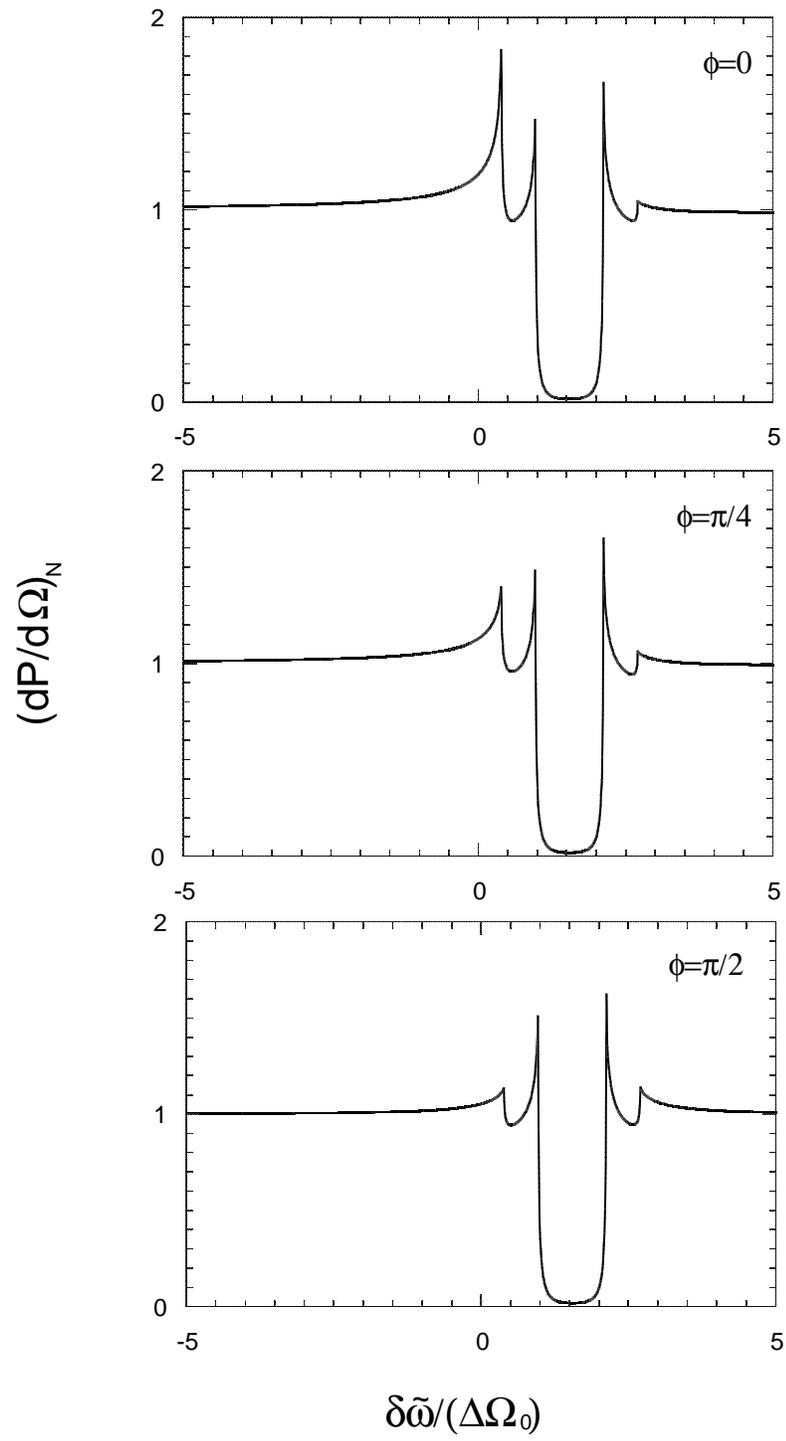
\begin{figure}
\caption{Normalized and averaged power emission $(dP/d\Omega)_N$ at $\theta=60^o$ for $\phi=0, \pi/4$ and $\pi/2$. The rest of the parameters are the same as in Fig. 2.}
\end{figure}

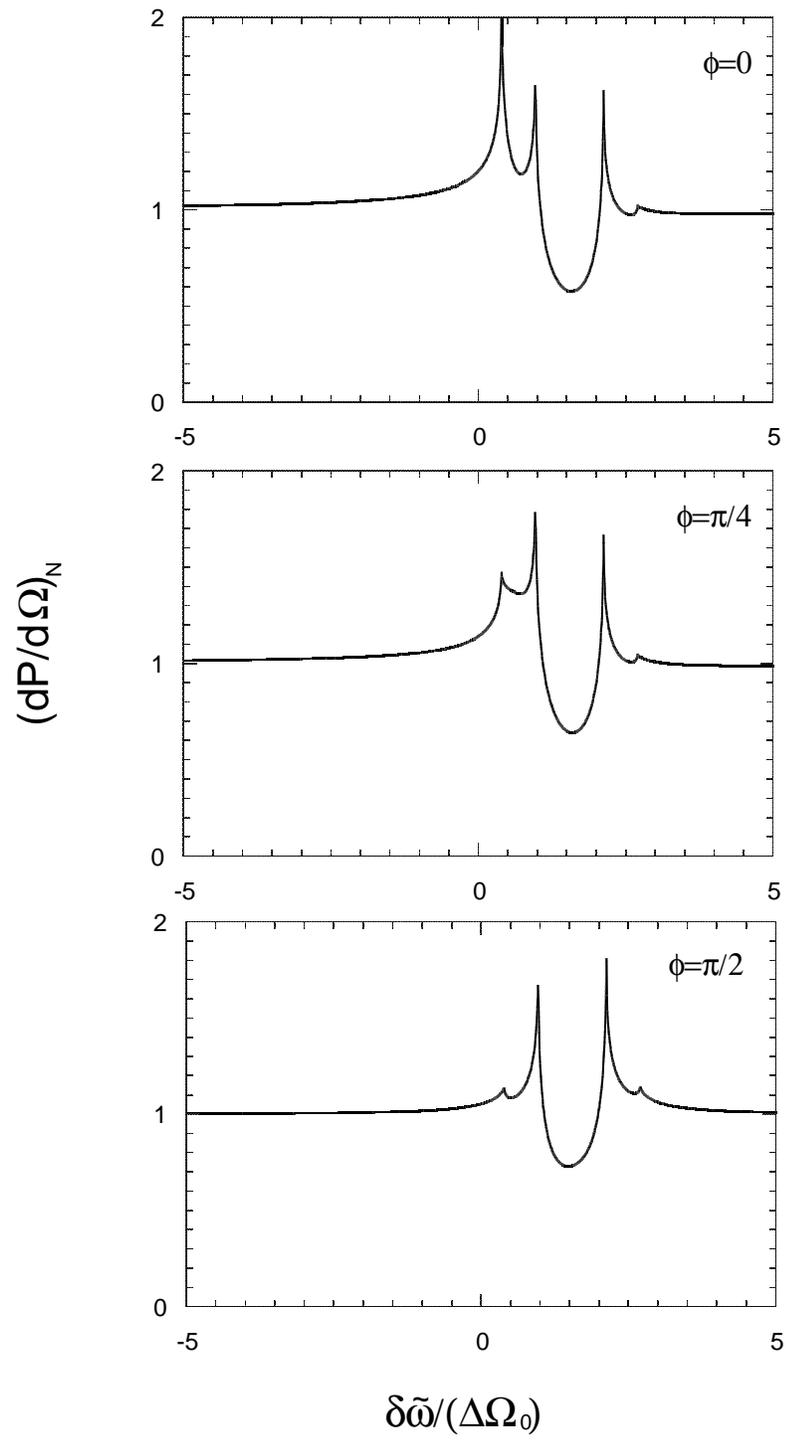
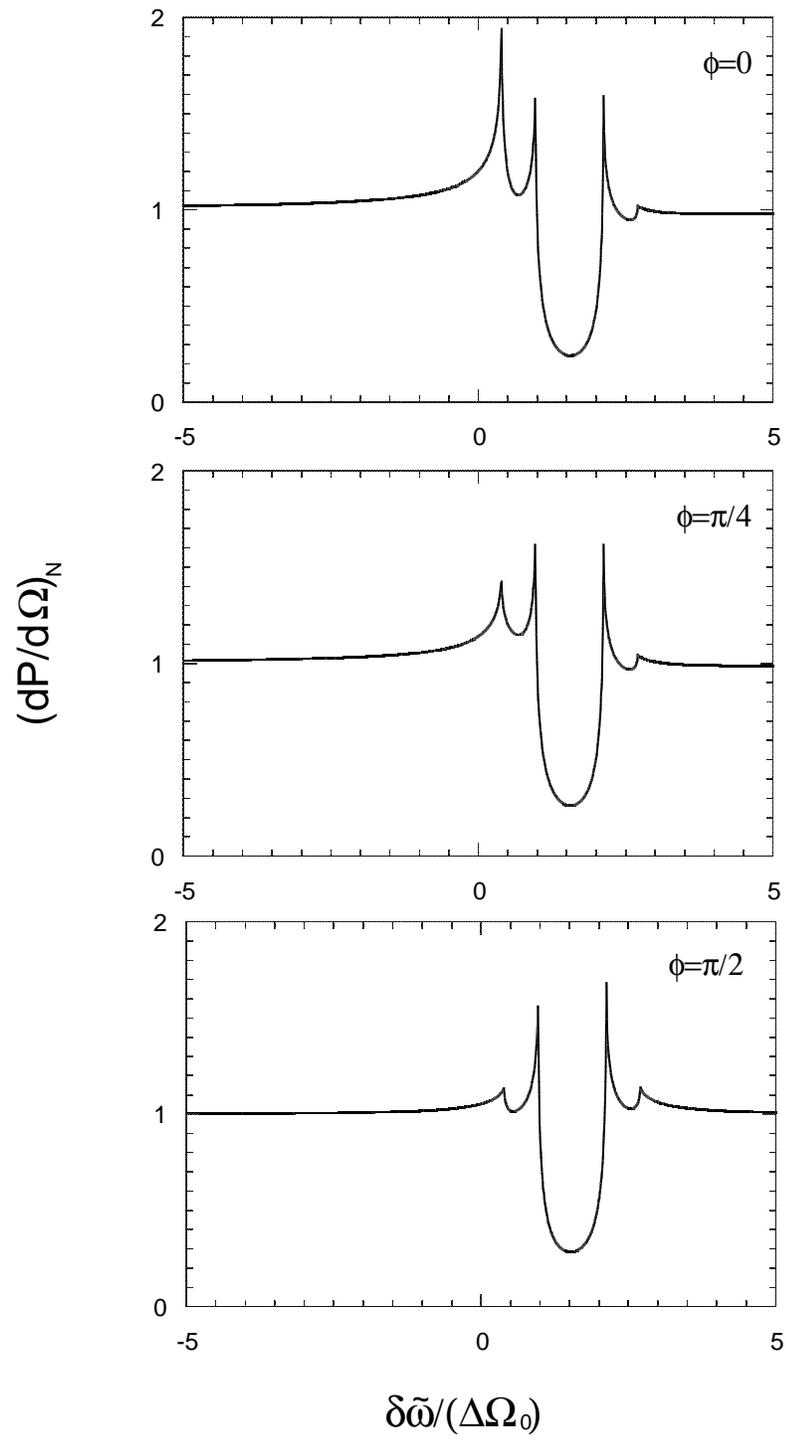
\begin{figure}
\caption{Emission spectrum of the emitter located at (a) $N=1$, (b) $N=2$  periods away from the boundary for $\theta=60^o$ and for the same set of the initial phases as in Fig. 3. Averaging over the position within the unit cell and over all possible orientaions of the dipole emitter direction has been performed.} 
\end{figure}

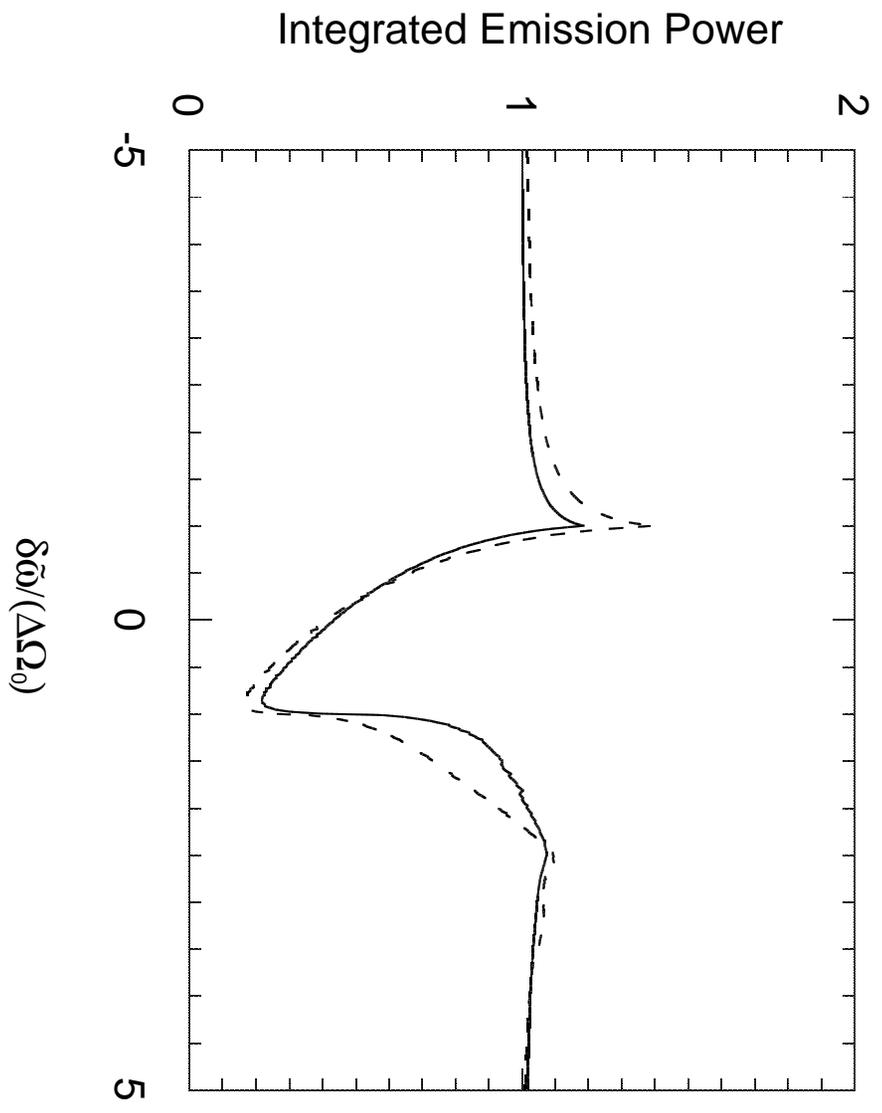
\begin{figure}
\caption{The total power emitted in $xy$ plane for  $\phi=0$ (solid line), $\phi=\pi$ (dashed line). Here $j_x=j_y=j_z$, $d=2a$. Averaging over the dipole emitter position within the unit cell has been performed.}
\end{figure}


\begin{references}
\bibitem{note}Plotting Eq. (\ref{power2}) for  different emitter--interface distances $d$ we realized that $d=5a$ differs only weakly from $d=\infty$, in the sense, that the typical value of the power intensity within the gap amounts to less than $5\%$ of its value outside the gap.
\end{references}
\end{document}